\documentclass[twocolumn,showpacs,preprintnumbers,superscriptaddress,amsmath,amssymb,prl]{revtex4}

\usepackage{graphicx}% Include figure files
\usepackage{dcolumn}% Align table columns on decimal point
\usepackage{bm}% bold math

\newcommand{\ipb}{\ensuremath{\mathrm{pb^{-1}}}}

\newcommand{\TeV}{\ensuremath{\mathrm{Te\kern -0.1em V}}}
\newcommand{\TeVc}{\ensuremath{\mathrm{Te\kern -0.1em V\!/}c}}
\newcommand{\TeVcc}{\ensuremath{\mathrm{Te\kern -0.1em V\!/}c^2}}
\newcommand{\GeV}{\ensuremath{\mathrm{Ge\kern -0.1em V}}}
\newcommand{\GeVc}{\ensuremath{\mathrm{Ge\kern -0.1em V\!/}c}}
\newcommand{\GeVcc}{\ensuremath{\mathrm{Ge\kern -0.1em V\!/}c^2}}
\newcommand{\MeV}{\ensuremath{\mathrm{Me\kern -0.1em V}}}
\newcommand{\MeVc}{\ensuremath{\mathrm{Me\kern -0.1em V\!/}c}}
\newcommand{\MeVcc}{\ensuremath{\mathrm{Me\kern -0.1em V\!/}c^2}}

\begin{document}
%==============================================================================

\title{
{\raggedleft\small
\vspace*{-1.0cm}
FERMILAB-PUB-05/535-E\\
%CDF/PUB/BOTTOM/CDFR/???\\
%(For submission to  for. Phys. Rev. Lett. --- \today)\\
}\vspace*{0.2cm}
\boldmath
Measurement of the Dipion Mass Spectrum in  $X(3872)\rightarrow J/\psi\,\pi^{+}\pi^{-}$ Decays
}

\affiliation{Institute of Physics, Academia Sinica, Taipei, Taiwan 11529, Republic of China} 
\affiliation{Argonne National Laboratory, Argonne, Illinois 60439} 
\affiliation{Institut de Fisica d'Altes Energies, Universitat Autonoma de Barcelona, E-08193, Bellaterra (Barcelona), Spain} 
\affiliation{Baylor University, Waco, Texas  76798} 
\affiliation{Istituto Nazionale di Fisica Nucleare, University of Bologna, I-40127 Bologna, Italy} 
\affiliation{Brandeis University, Waltham, Massachusetts 02254} 
\affiliation{University of California, Davis, Davis, California  95616} 
\affiliation{University of California, Los Angeles, Los Angeles, California  90024} 
\affiliation{University of California, San Diego, La Jolla, California  92093} 
\affiliation{University of California, Santa Barbara, Santa Barbara, California 93106} 
\affiliation{Instituto de Fisica de Cantabria, CSIC-University of Cantabria, 39005 Santander, Spain} 
\affiliation{Carnegie Mellon University, Pittsburgh, PA  15213} 
\affiliation{Enrico Fermi Institute, University of Chicago, Chicago, Illinois 60637} 
\affiliation{Joint Institute for Nuclear Research, RU-141980 Dubna, Russia} 
\affiliation{Duke University, Durham, North Carolina  27708} 
\affiliation{Fermi National Accelerator Laboratory, Batavia, Illinois 60510} 
\affiliation{University of Florida, Gainesville, Florida  32611} 
\affiliation{Laboratori Nazionali di Frascati, Istituto Nazionale di Fisica Nucleare, I-00044 Frascati, Italy} 
\affiliation{University of Geneva, CH-1211 Geneva 4, Switzerland} 
\affiliation{Glasgow University, Glasgow G12 8QQ, United Kingdom} 
\affiliation{Harvard University, Cambridge, Massachusetts 02138} 
\affiliation{Division of High Energy Physics, Department of Physics, University of Helsinki and Helsinki Institute of Physics, FIN-00014, Helsinki, Finland} 
\affiliation{University of Illinois, Urbana, Illinois 61801} 
\affiliation{The Johns Hopkins University, Baltimore, Maryland 21218} 
\affiliation{Institut f\"{u}r Experimentelle Kernphysik, Universit\"{a}t Karlsruhe, 76128 Karlsruhe, Germany} 
\affiliation{High Energy Accelerator Research Organization (KEK), Tsukuba, Ibaraki 305, Japan} 
\affiliation{Center for High Energy Physics: Kyungpook National University, Taegu 702-701; Seoul National University, Seoul 151-742; and SungKyunKwan University, Suwon 440-746; Korea} 
\affiliation{Ernest Orlando Lawrence Berkeley National Laboratory, Berkeley, California 94720} 
\affiliation{University of Liverpool, Liverpool L69 7ZE, United Kingdom} 
\affiliation{University College London, London WC1E 6BT, United Kingdom} 
\affiliation{Massachusetts Institute of Technology, Cambridge, Massachusetts  02139} 
\affiliation{Institute of Particle Physics: McGill University, Montr\'{e}al, Canada H3A~2T8; and University of Toronto, Toronto, Canada M5S~1A7} 
\affiliation{University of Michigan, Ann Arbor, Michigan 48109} 
\affiliation{Michigan State University, East Lansing, Michigan  48824} 
\affiliation{Institution for Theoretical and Experimental Physics, ITEP, Moscow 117259, Russia} 
\affiliation{University of New Mexico, Albuquerque, New Mexico 87131} 
\affiliation{Northwestern University, Evanston, Illinois  60208} 
\affiliation{The Ohio State University, Columbus, Ohio  43210} 
\affiliation{Okayama University, Okayama 700-8530, Japan} 
\affiliation{Osaka City University, Osaka 588, Japan} 
\affiliation{University of Oxford, Oxford OX1 3RH, United Kingdom} 
\affiliation{University of Padova, Istituto Nazionale di Fisica Nucleare, Sezione di Padova-Trento, I-35131 Padova, Italy} 
\affiliation{LPNHE-Universite de Paris 6/IN2P3-CNRS} 
\affiliation{University of Pennsylvania, Philadelphia, Pennsylvania 19104} 
\affiliation{Istituto Nazionale di Fisica Nucleare Pisa, Universities of Pisa, Siena and Scuola Normale Superiore, I-56127 Pisa, Italy} 
\affiliation{University of Pittsburgh, Pittsburgh, Pennsylvania 15260} 
\affiliation{Purdue University, West Lafayette, Indiana 47907} 
\affiliation{University of Rochester, Rochester, New York 14627} 
\affiliation{The Rockefeller University, New York, New York 10021} 
\affiliation{Istituto Nazionale di Fisica Nucleare, Sezione di Roma 1, University of Rome ``La Sapienza," I-00185 Roma, Italy} 
\affiliation{Rutgers University, Piscataway, New Jersey 08855} 
\affiliation{Texas A\&M University, College Station, Texas 77843} 
\affiliation{Istituto Nazionale di Fisica Nucleare, University of Trieste/\ Udine, Italy} 
\affiliation{University of Tsukuba, Tsukuba, Ibaraki 305, Japan} 
\affiliation{Tufts University, Medford, Massachusetts 02155} 
\affiliation{Waseda University, Tokyo 169, Japan} 
\affiliation{Wayne State University, Detroit, Michigan  48201} 
\affiliation{University of Wisconsin, Madison, Wisconsin 53706} 
\affiliation{Yale University, New Haven, Connecticut 06520} 
\author{A.~Abulencia}
\affiliation{University of Illinois, Urbana, Illinois 61801}
\author{D.~Acosta}
\affiliation{University of Florida, Gainesville, Florida  32611}
\author{J.~Adelman}
\affiliation{Enrico Fermi Institute, University of Chicago, Chicago, Illinois 60637}
\author{T.~Affolder}
\affiliation{University of California, Santa Barbara, Santa Barbara, California 93106}
\author{T.~Akimoto}
\affiliation{University of Tsukuba, Tsukuba, Ibaraki 305, Japan}
\author{M.G.~Albrow}
\affiliation{Fermi National Accelerator Laboratory, Batavia, Illinois 60510}
\author{D.~Ambrose}
\affiliation{Fermi National Accelerator Laboratory, Batavia, Illinois 60510}
\author{S.~Amerio}
\affiliation{University of Padova, Istituto Nazionale di Fisica Nucleare, Sezione di Padova-Trento, I-35131 Padova, Italy}
\author{D.~Amidei}
\affiliation{University of Michigan, Ann Arbor, Michigan 48109}
\author{A.~Anastassov}
\affiliation{Rutgers University, Piscataway, New Jersey 08855}
% ...
\collaboration{CDF Collaboration}
\noaffiliation

\date{16 December 2005 % \today\
}% It is always \today, today,
             %  but any date may be explicitly specified

\begin{abstract}
  We measure the dipion mass spectrum in  $X(3872)\rightarrow J/\psi\,\pi^+\pi^-$ decays
  using  $360~\ipb$  of $\bar{p} p $
  collisions at $\sqrt{s}=1.96~\TeV$ collected with the CDF II detector. 
  The spectrum is fit with   predictions for odd  $C$-parity 
 ($^3\!S_1$,  $^1\!P_1$, and $^3\!D_J$)
  charmonia decaying to  $J/\psi\, \pi^+\pi^-$,
  as well as even $C$-parity states in which the pions are from
    $\rho^0$ decay.
  The latter case also encompasses exotic interpretations, such as a $D^0\overline{D}{^{*0}}$ molecule.
  Only the  $^3\!S_1$   and  $J/\psi\,\rho$ hypotheses
  are compatible  with our data.
Since  $^3\!S_1$ is untenable on other grounds,  
decay via $J/\psi\,\rho$ is favored, which 
implies $C=+1$  for the   $X(3872)$.
Models for different  $J/\psi$-$\rho$ angular momenta $L$ are considered.
Flexibility in the models, especially the introduction of  
$\rho$-$\omega$ interference,
enable good  descriptions of our data for both $L=0$ and 1.
\end{abstract}

%   14.40.Gx, %  Properties of specific particles:  Mesons with S=C=B=0, mass>2.5 GeV (including quarkonia)
%   13.25.Gv,  %  Hadronic decays of mesons: Decays of J/psi, Upsilon, and other quarkonia 
%   12.39.Mk %  Glueball and nonstandard multi-quark/gluon states 
%    }     % PACS, the Physics and Astronomy
%                                % Classification Scheme.
%   %\keywords{Suggested keywords}%Use showkeys class option if keyword
%                                 %display desired
%   
%   

\maketitle

\begin{figure*}[t]
  \includegraphics[scale=0.93]{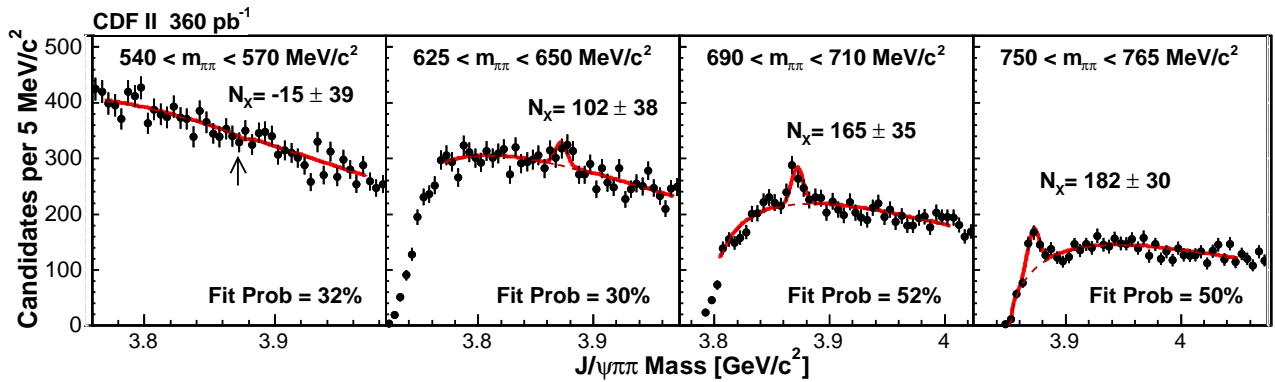}
  \caption{\label{slices} 
    Examples of ``slices''  in dipion mass $m_{\pi\pi}$ of the  $J/\psi\,\pi^{+}\pi^{-}$ mass distributions,
    and their  $X(3872)$ fits. 
    The raw yields $N_X$ prior to efficiency corrections are quoted,
    and the arrow in the first panel   marks   the $X(3872)$ mass.
  }
\end{figure*}

The charmonium-like $X(3872)$ stands as a major spectroscopic puzzle. 
Its mass~\cite{Belle3872,XCDF,XD0,BaBar3872mass}
and what is known of its decays make  assignments
to the normal spectrum of  $c\bar{c}$ states
problematic~\cite{ccbarOptions, BelleNewMass}.
Its remarkable proximity    
to the $D^0\overline{D}{^{*0}}$ 
mass---indistinguishable within uncertainties---has fueled
speculations that it is
a loosely bound deuteron-like $D^0\overline{D}{^{*0}}$ ``molecule'',
{\it i.e.} a $(u\bar{c})$-$(c\bar{u})$ system~\cite{Belle3872,SwansonMol}.
Although a molecule is prominent  among  exotic  interpretations,
others have been proffered~\cite{OtherExotics}.
Non-$q\bar{q}$ mesons are allowed within QCD,
but an unequivocal example remains elusive.
Even as a conventional meson, the  $X(3872)$ 
remains interesting as the  $c\bar{c}$ spectrum
above the $\psi(3770)$ is not well known.

Insight into the  $X(3872)$ is offered by
the dipion mass spectrum in  $X\rightarrow J/\psi\, \pi^{+} \pi^{-}$. 
Belle observed a preference for high $\pi\pi$-masses, 
contrary to expectations for 
triplet-$D$  $c\bar{c}$-states~\cite{Skwarnicki}---the na\"{\i}ve interpretation. 
Belle noted that $X\rightarrow J/\psi\,\rho^0$ decay---isospin violating 
for charm\-onium---produces high masses, and 
thus may be a hint for a $D^0\overline{D}{^{*0}}$ molecule.
Dipion spectra have been published~\cite{Belle3872,BaBar3872pipimass}; 
and a preliminary analysis partially based on $\pi\pi$ masses argues 
for a $J^{PC}=1^{++}$ assignment~\cite{BelleNewMass}, 
consistent with that expected for  a  $D^0\overline{D}{^{*0}}$ molecule.

At the Tevatron, large $X(3872)$ samples are available,
albeit with high backgrounds. 
Previously, we have  measured the $X$ mass
and confirmed the propensity for high $\pi\pi$ masses~\cite{XCDF}.
We also have made a preliminary measurement of the inclusive production fraction
arising from $b$ hadrons~\cite{cdfDPF04}.
Here we measure the $\pi\pi$ mass spectrum.

We use a  sample of $\bar{p}p$ collisions at $\sqrt{s}=1.96~\TeV$
collected  with the  Collider Detector at Fermilab (CDF~II)
between February 2002 and August 2004.
The detector is described in detail elsewhere~\cite{CDFgeneral},
and only the most relevant components for this analysis are summarized here.
The  central tracking system is immersed in a 1.4 T solenoidal magnetic field 
for the measurement of charged particle momenta  $p_T$ transverse to the beamline.
It is composed of six layers of silicon-strip detectors 
(L00~\cite{L00} and SVX~\cite{SVXII})  surrounded by an 
open-cell  drift chamber called the Central Outer Tracker  (COT)~\cite{COT}.   
The active volume of the COT is a 3.1 m long cylinder 
with 8~superlayers of 12~wires each.
The outermost detection system is planes of multi-layer drift 
chambers for detecting muons~\cite{Muons}.
The Central Muon system (CMU) covers
 $\left| \eta \right| \leq  0.6$,
where pseudorapidity $\eta \equiv -\ln[\tan(\theta/2)]$ and $\theta$ is the angle 
of the particle with respect to the direction of the proton beam.  
Additional chambers (CMX)  
extend the muon coverage to $\left| \eta \right| = 1.0$.

A dimuon trigger is used to obtain a $J/\psi\,\to\mu^+\mu^-$ sample.
At  Level 1 of a three-level trigger system,
the  eXtremely Fast Tracker (XFT)~\cite{XFT} uses COT information 
to select tracks based on $p_{T}$.  
XFT tracks with  $p_T \geq 1.5$ $(2.0)~\GeVc$ are
extrapolated to the CMU (CMX) chambers and compared with 
the positions of muon-chamber tracks.  
The event passes Level 1 if
two or more XFT tracks 
are matched to muon tracks.
Opposite-charge and opening-angle cuts 
are imposed at Level~2. 
At Level~3, full COT tracking information
is used to reconstruct $\mu^+\mu^-$ candidates.
Events with candidates 
from $2.7$ to  $4.0~\GeVcc$ in mass
are recorded for further analysis.

This analysis~\cite{Sasha} is based on  an integrated luminosity of $360~\ipb$.
Candidate selection follows Ref.~\cite{XCDF} with two exceptions (see below).
After constraining $\mu^+\mu^-$ candidates to a common 
vertex,  the  dimuon mass must be  within $60~\MeVcc$ 
($\sim\!4$ standard deviations) 
of the $J/\psi$ mass~\cite{PDG}.
This  one degree-of-freedom ($DoF$) fit  must have   $\chi^2<15$.
Pairs of charged tracks, each with $p_T \geq 0.4~\GeVc$ and assumed to be pions,
are fit with the $\mu^+\mu^-$ tracks to a common vertex.
In this fit,  the dimuon mass is
constrained to the $J/\psi$ mass, 
and  we demand $\chi^2<25$ (6 $DoF$).
We reduce combinatorial backgrounds by requiring
$p_T(J/\psi) \geq 4~\GeVc$
and $\Delta R \leq 0.7$ for both pions, 
where $\Delta R  \equiv \sqrt{ (\Delta \phi)^{2} + (\Delta \eta)^{2} }$,  $\Delta \phi$ 
is the difference in azimuthal angle between the $J/\psi\, \pi^{} \pi^{}$ system
and the pion, and 
$\Delta \eta$ is the difference in pseudorapidity.
The mass range for the sample includes both 
$X(3872)$ and $\psi(2S)$ signals~\cite{XCDF};
the latter is used as a control sample.

We depart from Ref.~\cite{XCDF} by dropping a cut
on the number of candidates allowed per event.
This removes a possible bias and  improves
the  $X$ signal at high $\pi\pi$ masses.
We also add fiducial criteria: 
$p_T(J/\psi\, \pi^{} \pi^{}) > 6\,\GeVc$ and $|\eta(J/\psi\, \pi^{} \pi^{})|<0.6$.
This eliminates the region of  rapidly changing efficiency
and sacrifices $\sim\!25\%$ of the $\psi(2S)$ yield, leaving
$11500\pm 220$  $\psi(2S)$ mesons.
We have $1260\pm 130$  $X(3872)$ candidates
for $m_{\pi\pi}>500\,\MeVcc$~\cite{XCDF}.

To extract $dN/dm_{\pi\pi}$ spectra,
we divide the sample into ``slices'' of $m_{\pi\pi}$ and fit 
each  $J/\psi\, \pi^{+} \pi^{-}$ mass distribution for 
the  signal per slice.
The  $J/\psi\, \pi^{+} \pi^{-}$ mass fits use
a Gaussian for the $X(3872)$ signal
and an exponential times power law for the background.
We also fit the $\psi(2S)$ control signal in the same way,
but use two Gaussians for the better defined shape
arising from the larger  $\psi(2S)$ signal. 
As slices may have small signals---or none at all---we
inhibit the fit from latching onto fluctuations by fixing the position and
width of the signal to values  from  full-sample fits.
Sample slices are shown in Fig.~\ref{slices}.

The fitted $dN/dm_{\pi\pi}$ yield is corrected for detector and selection
efficiencies determined by Monte Carlo simulation.
Only the efficiencies  relative 
to other  $m_{\pi\pi}$ slices are needed. 
An important input affecting the  efficiency is the 
production $p_T$ spectrum.
For  the  $X(3872)$,
there is no {\it a priori} model,
and we  rely upon data.
The generated spectra, exponentials raised to a quadratic polynomial in $p_T$, 
are adjusted until the simulation, after
detector and reconstruction effects,  
reproduces the respective  $X(3872)$  or $\psi(2S)$  
spectra of the data.
We use a  parameterization of the  well-known  $m_{\pi\pi}$  shape
of the $\psi(2S)$ for both states.

Uncertainties on our spectra are dominated 
by statistics, but we examined two 
sources of  systematic effects: 
the fits for signal yields
and the efficiency corrections.

To check the  yield  stability,  we  changed  the 
width of the $J/\psi\,\pi^+\pi^-$  fit range of
$200~\MeVcc$ by $\pm50~\MeVcc$,
altered the signal and background models, and
allowed the signal mean and width to float.
We saw no bias in the yields, 
but nevertheless allotted an uncertainty based on the statistical
precision to which we could observe
one: 3.6 $\psi(2S)$  and 8.4 $X(3872)$ candidates per slice.
The highest three $X$-slices---highest two for the $\psi(2S)$---are
treated specially  for effects near the upper  kinematic limit:
the  background begins to turn-on  under the signal, and 
resolution effects can distort the signal shape in the %$J/\psi\, \pi^{+} \pi^{-}$ 
mass fit.
Yield systematics are assigned to these slices
based on variations in the fit model for these issues.

The other type of uncertainty is from the efficiency corrections.
The  $\psi(2S)$ model for  $dN/dm_{\pi\pi}$  is inexact. 
We assign an uncertainty based on  phase space as an alternate
shape---including re-tuning  the $p_T$ spectrum. 
The ratio of the alternate correction  to the nominal one 
quantifies the change in  {\it shape}
of the efficiency  when switching from the  $\psi(2S)$-like
 $dN/dm_{\pi\pi}$
% $m_{\pi\pi}$ 
to phase space.
The  ratio of efficiencies for the $\psi(2S)$  varies  by $\sim3\%$
over the main region of interest, $m_{\pi\pi}>360~\,\MeVcc$,
and by $\sim2.5\%$ for  the $X(3872)$ 
above $570~\,\MeVcc$.
For the  uncertainty in the meson  $p_T$ spectra,
we use alternate spectra one standard deviation
steeper and shallower in their $p_T$ fall-off
based on the errors from the  $p_T$ spectrum fit to the data.
We  again take the ratio of the new efficiencies  relative to the nominal shape 
to  quantify the uncertainty.
For the mass ranges of interest the  $\psi(2S)$ variation is almost $3\%$,
but less than $1\%$ for the  $X$.
The  $p_T$ spectrum of the $X$ is more poorly measured than for the  $\psi(2S)$;
but with  higher dipion masses,
the $X$ suffers smaller variations in efficiency, 
and thus a smaller uncertainty.

  \begin{figure}[t]
  \includegraphics[scale=0.44]{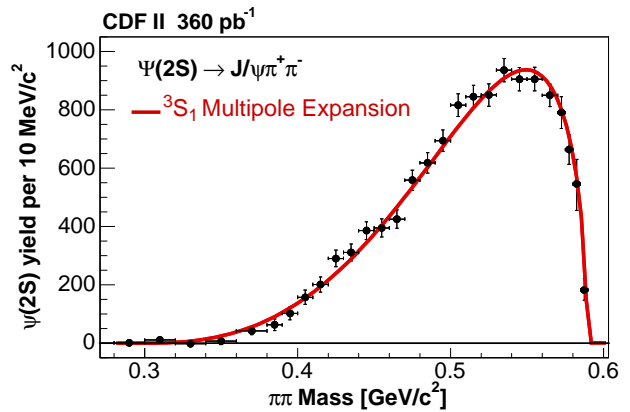}
  \caption{\label{PsiMpipi} 
  The CDF dipion mass  spectrum for the $\psi(2S)$ with a fit of a 
  QCD multipole expansion calculation for $^3\!S_1$~\cite{Yan80}.
  }
  \end{figure}

The $dN/dm_{\pi\pi}$ spectrum for our  $\psi(2S)$ control signal,
after corrections and including  systematic uncertainties, 
is shown  in Fig.~\ref{PsiMpipi}
with a  scale  preserving the raw fitted yield of $11\,500$ candidates.
It agrees well with results from the BES Collaboration 
using a sample of $20\,000$ events~\cite{BES}.
This is reflected by the mutual agreement in fits
to  a QCD  multipole expansion model~\cite{Yan80}.
BES obtained  $-0.336 \pm 0.009 \pm 0.019$
for this model's single shape parameter ``$B/A$,'' 
whereas our fit yields  $-0.342\pm0.022$  (6.9\% fit probability).
The systematic  uncertainties are incorporated in this
and later $X$-fits, 
including the  $m_{\pi\pi}$  correlations in the  efficiency uncertainties.

\begin{figure}[t]
\includegraphics[scale=0.44]{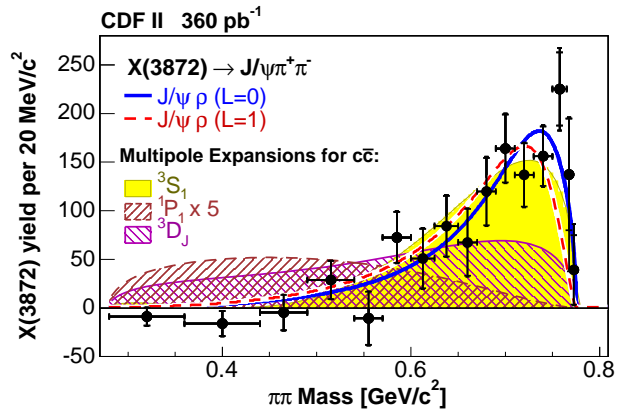}
\caption{\label{MpipiX} 
The dipion mass  spectrum for the $X(3872)$, and fits
to various hypotheses (see text).
The fitted curve for the  $^1\!P_1$ model is scaled up by a factor of 5 for
better visibility.
}
\end{figure}

The  $X(3872)$ dipion  spectrum 
is shown in Fig.~\ref{MpipiX}.
We fit our data with multipole expansion 
calculations for $C$-odd  $c\bar{c}$ options~\cite{MEcaveat}.
The  $^3\!D_J$ states are  a natural choice for the  $X$
according to potential models~\cite{ccbarOptions}.
A fit of   $^3D_J\rightarrow J/\psi\, \pi^+\pi^-$~\cite{Yan80}
is unacceptable 
with a $\chi^2$ of 113 for 14 $DoF$.
The  $^1P_1 \rightarrow J/\psi\, \pi^+\pi^-\,$\cite{1P1}
fit is worse ($\chi^2/DoF=$ $146/14$).
The $\psi(2S)$  
spectrum is similar  to  that of the  $X$, and  
indeed, our  $^3S_1$~\cite{Yan80} fit to the $X$ has
a $28  \%$ probability. 
However, 
no  new $^3S_1$ $c\bar{c}$-state can be near $3872\,\MeVcc$
as the $\psi(3S)$ 
% $3\,^3S_1$ 
lies  at $\sim\!4040~\MeVcc$~\cite{ccbarOptions}.

The above $C$-odd  states produce  dipions,
to lowest $L$ between the pions,
with $J^{PC}=0^{++}$.
$C$-even  states yield  $1^{--}$ isovector  dipions, which we associate 
with the $\rho^0$. 
Isospin conservation suppresses  $c\bar{c}$ decays  to $J/\psi\,\rho$.
Thus, this  mode is seen as  suggestive of a $D^0\overline{D}{^{*0}}$ molecule~\cite{Belle3872,SwansonMol}.
Even as charmonium, however, the $X$ may break isospin
by coupling to  $D^0\overline{D}{^{*0}}$
due to its close  proximity in mass.

We  model  $X \rightarrow J/\psi\, \rho$ by
a relativistic  Breit-Wigner 
multiplied by  phase space 
and generalized  for  a  $J/\psi$-$\rho$ system of
angular momentum  $L$.
That is, 
$dN/dm_{\pi\pi} 
\propto$ $k_*^{2{L}+1} f^2_{LX}(k_*) |B_\rho|^2$,
where
$k_*$ is the $J/\psi$  momentum in the $X$ rest-frame,
$B_\rho \propto  \sqrt{ m_{\pi\pi}    % m_{\rho} 
\Gamma_\rho}/[  m_\rho^2 - m_{\pi\pi}^2   - i m_\rho\Gamma_\rho ]$, 
$\Gamma_\rho(m_{\pi\pi}) = 
\Gamma_0      \left[ q_*/q_0 \right]^3      % \,\times$ $
                     [m_\rho /m_{\pi\pi}]
                     [f_{1\rho}(q_*)/f_{1\rho}(q_0)]^2 $,
$q_*$ is  the $\pi$ momentum in the $\rho$ rest-frame,
$q_0 \equiv q_*(m_\rho)$,
$ m_\rho$ is $775.8~\MeVcc$, and
$\Gamma_0$ is $146.4~\MeV$.
The $f_{Li}(p)$  are
$f_{0i}(p) = 1$  and 
$f_{1i}(p) = (1+R^2_i p^2)^{-\frac{1}{2}}$~\cite{BlattWeiss},
where  $R_i$ is 
a radius of interaction
for meson ``$i$.''
The $R_i$ are poorly known. 
A common value for light mesons is  $0.3$ fm,
but for $D$ mesons larger values like  1 fm  are often taken~\cite{Rstuff}.
We use these respective values  for $R_{\rho}$ and  $R_X$.

Fits with this $\rho$ model
are shown in  Fig.~\ref{MpipiX} for $L=0$ and 1.
Higher $L$ softens the fall-off at the high kinematic limit,
worsening the agreement:
the fit probability  goes from $55\%$ for  $L=0$
down to $7.7\%$ for  $L=1$~\cite{BelleLFits}.
The $P$-wave fit is somewhat disfavored, but the results are
sensitive to  $R_X$ and $R_\rho$.
The  latter probability can be increased by lowering 
$R_\rho$ or raising  $R_X$.

Another modeling uncertainty is 
the effect 
of  $\rho$-$\omega$ interference.
Belle reports evidence for  $X\rightarrow J/\psi\, \pi^+\pi^-\pi^0$
and interprets it as decay via a virtual $\omega$.
As such, they 
measure the ratio of 
$ J/\psi\,\omega$ to $ J/\psi\,\rho$
branching ratios ${\cal R}_{3/2}$
to be $1.0 \pm 0.4 \pm 0.3$~\cite{BellePsiGammaOmega}.
The rate of $\omega \rightarrow \pi^+\pi^-$ is  normally negligible, % here, 
but its interference effects may not be.

\begin{figure}[t]
\includegraphics[scale=0.455]{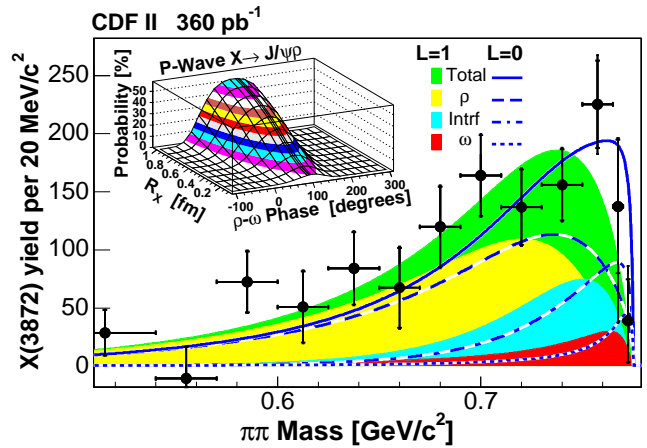}
\caption{\label{MpipiInterf} 
A blow-up of the  $X(3872)$ spectrum  with  $J/\psi\,\rho$ fits
which include $\rho$-$\omega$ interference ($95^\circ$ phase)
with  relative amplitudes set by ${\cal R}_{3/2}=1.0$.
Fits for both  $L=0$ (lines) and 1 (shaded regions) are shown, 
along with their decomposition into $\rho$, $\omega$, and interference terms.
The inset shows $L=1$ fit probabilities as a function of $\phi$ and $R_X$
in  5\% contours.
}
\end{figure}

We generalize  $|B_\rho|^2$ to $|A_\rho B_\rho + e^{i\phi}A_\omega B_{\omega2\pi}|^2$
in $dN_{2\pi}/dm_{\pi\pi}$,
where $A_\rho$ and $A_\omega$ are (positive)  decay amplitudes
via $\rho$ and $\omega$, and $\phi$ is their relative phase.
Using   $dN_{3\pi}/dm_{3\pi} \propto  |A_\omega B_{\omega3\pi}|^2$ 
for $J/\psi\, \pi^+\pi^-\pi^0$~\cite{RhoOmegDetails},
${\cal R}_{3/2}$ determines  $|A_\omega/A_\rho|$  given $\phi$.
We take a $\phi$ of $95^\circ$, 
the value if the only phase
is from  $\omega\rightarrow \pi^+\pi^-$  decaying via  $\rho$-$\omega$ mixing~\cite{ExpectPhi}.
Similar phases are seen in   $e^+e^- \rightarrow \pi^+\pi^-$~\cite{PhiMeasure}.
The $\omega$ fraction is small ($<10\%$), but % the 
interference is  constructive and contributes $\sim\!23$\% for both $L$, 
preferentially at high masses.
Fits with this model
are  shown in Fig.~\ref{MpipiInterf}, along with the breakdown 
into interference and
``pure'' $\rho$ and $\omega$ parts.
The  probability is 19\% for the $S$ fit,
and  53\% for the $P$ fit.
The results are not critically dependent on  ${\cal R}_{3/2}$:
probabilities remain above $7\%$ 
over a $\pm1$  standard deviation span
of Belle's ${\cal R}_{3/2}$ for both  $L$ values.
The $P$ fit is sensitive to  $\phi$ and $R_X$,
as is shown in the inset. 
We conclude that  there is ample flexibility
in models  of $X\rightarrow J/\psi\, \rho$ of  either  $L$ 
to describe our  data.

In summary, we measured the dipion mass spectrum in  
$X(3872) \rightarrow J/\psi\,\pi^+\pi^-$.
Our spectrum is  inconsistent with calculations for  $^1P_1$ and  $^3D_J$ charmonia.
A good fit is obtained for $X\rightarrow  J/\psi\,\rho^0$, 
an interpretation 
supported by recent evidence for   
the $C$-even decay
$X\rightarrow J/\psi\,\gamma$~\cite{BellePsiGammaOmega}. 
Our data are  compatible with
both  $S$- and $P$-wave $J/\psi\,\rho$ decays,
where in the latter case, this is partly due to 
modeling  uncertainties.
The $P$  fit  benefits from  constructive $\rho$-$\omega$ interference
at levels implied by the rate of $X\rightarrow J/\psi\,\pi^+\pi^-\pi^0$.
The $J/\psi\,\rho$ interpretation does not by itself
distinguish between $C$-even charmonia  (e.g. $1^{++}$ or $2^{-+}$) 
and exotic options
like  an  $1^{++}$  $D^0\overline{D}{^{*0}}$ molecule.

We  thank
E.~Eichten,
S.~Olsen,
P.~Ko,
Y.-P.~Kuang, and
C.~Quigg
for  helpful discussions,
and the Fermilab staff and the technical staffs of the participating institutions 
for their vital contributions. 
This work was supported by the U.S. Department of Energy and National Science Foundation; 
the Italian Istituto Nazionale di Fisica Nucleare; 
the Ministry of Education, Culture, Sports, Science and Technology of Japan; 
the Natural Sciences and Engineering Research Council of Canada; 
the National Science Council of the Republic of China; 
the Swiss National Science Foundation; 
the A.P. Sloan Foundation; the Bundesministerium f\"ur Bildung und Forschung, Germany; 
the Korean Science and Engineering Foundation and the Korean Research Foundation; 
the Particle Physics and Astronomy Research Council and the Royal Society, UK; 
the Russian Foundation for Basic Research; 
the Comisi\'on Interministerial de Ciencia y Tecnolog\'{\i}a, Spain; 
in part by the European Community's Human Potential Programme under contract HPRN-CT-2002-00292; 
and the Academy of Finland.

%==============================================================================

%==============================================================================

\begin{thebibliography}{99}
\bibitem{BelleLFits} 
Belle's $\rho$-fits in Ref.~\cite{BelleNewMass} do not include
Blatt-Weiss\-kopf $f_{Li}$ factors.
Fitting CDF data with their Breit-Wigner model
returns a $P$-wave probability  of $0.1\%$.
\end{thebibliography}
\end{document}